\documentclass[prb,notitlepage]{revtex4-1} 

\usepackage{amsmath} 
\usepackage{amsfonts}
\usepackage{graphicx} 
\usepackage{scalerel}[2016/12/29]

\usepackage[polish,english]{babel}
\usepackage[T1]{fontenc}

\begin{document}
\title{Gravitational time dilation, free fall, and matter waves}

\author{Anna P.~Czarnecka}
\email{atpruscz@ualberta.ca}
\affiliation{Department of Physics, University of Alberta, Edmonton,  Alberta, Canada T6G 2E1}
\author{Andrzej Czarnecki}
\email{andrzejc@ualberta.ca}
\affiliation{Department of Physics, University of Alberta, Edmonton,  Alberta, Canada T6G 2E1}

\begin{abstract}
We demonstrate that the de Broglie wave of a particle in a 
gravitational field turns towards the region of lower
gravitational potential, causing the particle to fall. This turning is
caused by clocks running slower in the smaller potential. We use the
analogy of ocean waves that are slower in shallower water and turn
towards beaches. This approach  implies that the motion is along a
geodesic and explains the free fall qualitatively
and quantitatively
with only elementary algebra. 
\end{abstract}
\maketitle
\section{Introduction}

Bodies fall because  matter waves refract due to the gravitational
time dilation --- this is the new interpretation we present here. 
That the free fall is caused by the gravitational time dilation has
been demonstrated, in a more complicated way, in a beautiful paper by
Roy Gould.\cite{gould:2016aa} 
He points out that although Einstein's model of gravity predicts distortions
of both time and space near massive bodies, ordinary objects travel
primarily through time and their motion is mainly influenced by
time dilation. In a region of lower gravitational potential, time
flows more slowly. 

Before continuing, we should make this statement more precise. An
observer far away from massive bodies assigns positions and times
to all events. From the point of view of this observer, a clock
placed near a massive body runs slow, and this effect is more
pronounced for clocks in a smaller (more negative) gravitational potential. 

For example, two clocks near Earth's surface, separated
by height $h$, tick at different rates. If the average  time
measured by them is $t$, the upper clock measures more  time by
$\Delta t$ (see Section \ref{dilat}),
\begin{equation}
\Delta t=\frac{gh}{c^{2}}t ,\label{DeltaT}
\end{equation}
where $g\simeq9.8\,{\text{m}}/{\text{s}^{2}}$ is the gravitational
acceleration and $c\simeq3\cdot10^{8}\,{\text{m}}/{\text{s}}$ is
the speed of light. 

This phenomenon is so important  that it has rightly become the
subject of stories for young children.\cite{greene2008icarus} In
Ref.~\onlinecite{gould:2016aa} it is qualitatively explained with the analogy
to air travel along great circles, contrasted with straight lines
on maps. For a quantitative description of the trajectory, the Schwarzschild
metric is applied to determine the shape of the geodesic. This
analysis is further developed in Ref.~\onlinecite{Rebilas2017}.

\begin{figure}[h]
\centering\includegraphics[scale=0.4]{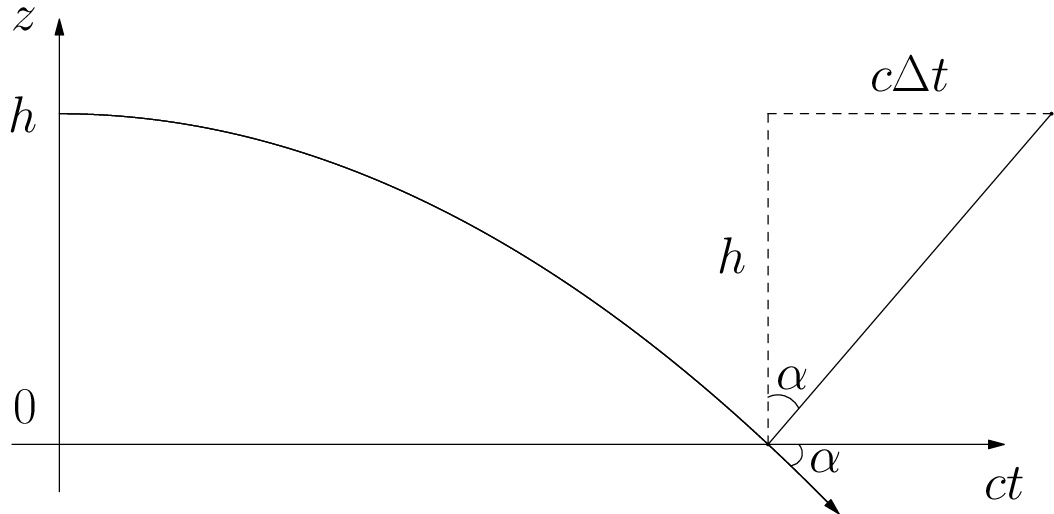}
\caption{Trajectory of a freely falling particle in spacetime (curved
  line). The de Broglie wave front (tilted solid line) changes
  direction in spacetime because time flows faster at a higher
  altitude.  The horizontal dashed line, labeled $c \Delta t$, is the
  extra displacement of the de Broglie wave at $z=h$ in the time
  dimension relative to the displacement in time at $z=0$. The tilted
  arrow indicates the velocity which, in spacetime, is a direction
  (characterized by the ratio $\text{d} z : \text{d} ct$).  This
  refraction is analogous to ocean waves turning towards a beach: they
  travel slower in shallower water.\label{deBroglie}}
\end{figure}

We propose a simpler approach using de Broglie matter waves, as
illustrated in 
Figure \ref{deBroglie}. We shall present our argument in terms of wave 
packets in Section \ref{packet}. Here we explain its gist.

Start with a particle initially at rest at a height $z=h$ above Earth's surface, with
$h$ small so that the gravitational acceleration can be assumed to be
constant in the range $0\le z \le h$. Throughout this paper we are
interested only in the motion in the vertical $z$ direction. Also, we
assume that the time flow rate differences are tiny  and make
corresponding approximations of the type $(1+\delta)^{-1} \simeq
1-\delta$. 

If there were no gravity, the particle would remain at rest at
$z=h$. In spacetime, the particle would move only through time, with
its de Broglie wave oscillating as
$\exp(-i mc^2 t /\hbar)$,\cite{Dieks:1990aa} where $m$ denotes the
particle's mass.

Consider now the effect of gravity on the de Broglie wave (see
Figure \ref{deBroglie}). Since the local time flows faster at larger
altitudes, de Broglie wave tilts and the particle starts moving
through space towards smaller values of $z$ (it falls). 

The slope of the trajectory plotted in Figure \ref{deBroglie} is
$\tan\alpha = -{{\text{d}} h}/{ {\text{d}} ( ct ) } = {v}/{ c } $
where $v$ is the vertical speed of the 
particle. Because the velocity is perpendicular to the wave
front,\cite{Houchmandzadeh:2020aa}  $\alpha$ is also the angle
between the wave front and the vertical, as shown in
Fig.~\ref{deBroglie},
 and $\tan\alpha$ is the ratio
of $c$ times the extra time elapsed at $h$ to the distance $h$, thus
\begin{equation}
  \label{eq:7}
 {v\over c} =  {c\Delta t \over h }  =  {cght \over c^2h }
 \Rightarrow v = gt. 
\end{equation}
The tilting (refraction) of the de Broglie wave reproduces the free fall
kinematics.

The diagram in Figure \ref{deBroglie} is inspired by Figure 6 in
Ref.~\onlinecite{Stannard_2016}. That very pedagogical paper does not,
however, use de~Broglie waves; its reasoning is classical, based on
Ref.~\onlinecite{gould:2016aa}. In that approach, the trajectory
follows from the postulate that the particle moves on a geodesic.

In our approach the refraction of the de Broglie wave naturally
determines the trajectory.\cite{Houchmandzadeh:2020aa} 
Since time flows slower closer to Earth's surface, de Broglie waves
evolve more slowly there and their front turns towards the surface, just
like the ocean waves turn towards a beach because they propagate
more slowly in shallower water. 
We do not need to use the notions of a
metric tensor or of a
geodesic, let alone calculate its shape, creating
significant conceptual, technical, and pedagogical simplification.

Eq.~\eqref{DeltaT} is derived in Section \ref{dilat}. Section
\ref{packet} repeats the above discussion of de Broglie wave
refraction slightly more rigorously, using wave packets.
Application of Eq.~\eqref{DeltaT} to the twin paradox is described in Section~\ref{other}.
We conclude in Section
\ref{summary}.  Appendix \ref{AppEin}
summarizes Einstein's 1907 derivation of Eq.~\eqref{DeltaT}.  
In Appendix \ref{waveeq} we determine the momentum evolution of a
wave packet by examining the non-relativistic limit of the Klein-Gordon equation.

A companion 3-minute film\cite{AniaFilm} presents
the main idea and explains how the gravitational time dilation
causes bodies to fall.

\section{Gravitational time dilation} \label{dilat}
\subsection{Gravitational time dilation from red shift}
In this Section we derive Eq.~\eqref{DeltaT} by considering the energy
a photon gains when falling in a gravitational field. Einstein's
original derivation using the Lorentz transformation is summarized in
Appendix \ref{AppEin}.

Consider a model clock consisting of a charged harmonic oscillator with
frequency $\nu_1$, placed in the Earth's gravitational field, at
 point 1 where the gravitational potential is $V_1$. When photons
emitted by the oscillating charge arrive at another point 2 with
the gravitational potential $V_2$, conservation of energy requires
their frequency to change to $\nu_2$,
\begin{equation}
  \label{eq:1}
    \nu_1 \left( 1 + {V_1 \over c^2}\right) 
= \nu_2 \left( 1 +  {V_2\over c^2}  \right).
\end{equation}
All processes occurring at point 1 with time intervals $\Delta t_1$ are
observed from point 2 at intervals $\Delta t_2$ such that, according to
Eq.~\eqref{eq:1},
\begin{equation}
  \label{eq:2}
  {\Delta t_2 \over \Delta t_1} = {\nu_1 \over \nu_2}  = { 1 +  {V_2\over c^2}
    \over 1 +  {V_1\over c^2} }.
\end{equation}
Near Earth's surface $V_i = gh_i$. Assuming $h_1 = 0$, $h_2  = h$,
\begin{equation}
  \label{eq:3}
   {\Delta t_2 \over \Delta t_1} 
  =  1 +  {gh\over c^2}.
\end{equation}
Denoting $\Delta t_2 - \Delta t_1 = \Delta t$ and $\Delta t_1 = t$ we
reproduce  Eq.~\eqref{DeltaT}, $\Delta t={ght}/{c^{2}}$.

\subsection{Experimental verification of $\Delta t$}
Many experiments have  demonstrated
relativistic effects on clocks, including recently with optical lattice
clocks on the Tokyo Skytree tower.\cite{Takamoto:2020aa} Clocks
placed on airplanes,\cite{Hafele166,Hafele168} rockets,\cite{GravityProbeA} and
satellites\cite{Delva_2018,Herrmann_2018} have also been used.

Especially valuable from a pedagogical point of view is Project
GREAT,  conducted by Tom Van Baak and his family; it is exceptionally well
documented.\cite{Van-Baak:2007aa}

Van Baak purchased three surplus portable cesium atomic clocks on eBay
and converted his minivan into a mobile time laboratory.  Before the
clocks were taken to a higher altitude, their readings were compared
against reference atomic clocks. After three days, the portable clocks
were transported by car by the Van Baak family 1340 meters up Mount
Rainier.  Measurements were collected for 40 hours.

\begin{figure}[h]
\centering\includegraphics[scale=0.4]{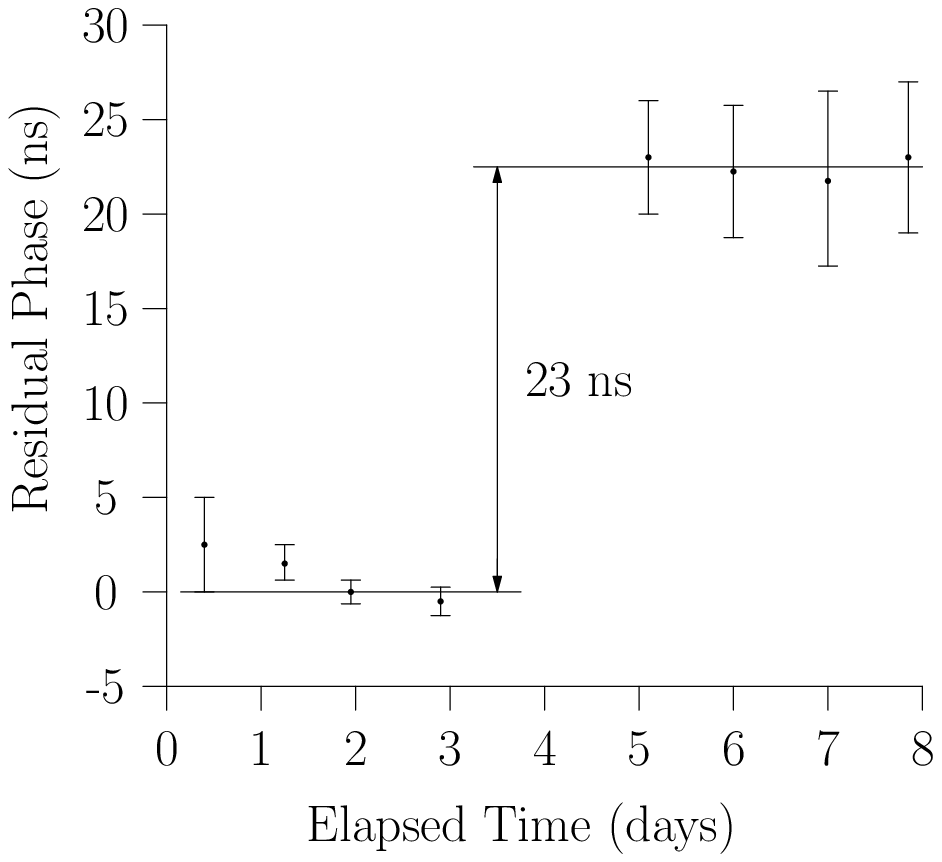}
\caption{Average readings of the atomic clock ensemble taken on the
  trip minus the readings of the reference
  clocks.\cite{Van-Baak:2007aa} The discontinuity between values
  before and after the mountain trip confirms gravitational time
  dilation, Eq.~\eqref{DeltaT}.\label{fig:great}}
\end{figure}

After returning, the clocks ran for another three days while being
compared with reference clocks that had remained at ground level. The
average extra time counted in the three clocks while up on the
mountain for two days was 23 nanoseconds (see Figure \ref{fig:great}),
in good agreement with Eq.~\eqref{DeltaT}, 
\begin{equation}
  \label{eq:6}
  \Delta t=\frac{gh}{c^{2}}t = {9.8\cdot 1340 \over 9\cdot 
    10^{16}}40\cdot 3600 \text{ s} =  21\text{ ns}. 
\end{equation}

\section{Wave packet near Earth }\label{packet}

Suppose that at $t=0$ the wave function of a particle has a Gaussian
shape in $z$, of width $\sigma$ and center at $z=0$,
\begin{equation}
\psi\left(z,t=0\right)=\frac{1}{\sqrt{\sigma\sqrt{\pi}}}
                                  \exp-\frac{z^{2}}{2\sigma^{2}}.
\end{equation}
Decompose $\psi$ into normalized momentum eigenstates,
\begin{equation}
\phi\left(k\right)  =\int_{-\infty}^{\infty}{\text{d}} z\psi\left(z\right)e^{-ikz}=\sqrt{2\sigma\sqrt{\pi}}e^{-\frac{k^{2}\sigma^{2}}{2}}.
\end{equation}
A state with a wave vector $k$ has the energy
$E_{k}=c\sqrt{m^{2}c^{2}+\hbar^{2}k^{2}}\simeq mc^{2}+ {\hbar^{2}k^{2}}/{2m}$. The wave function
evolves as (see Appendix \ref{waveeq})
\begin{equation}
\psi\left(z,t^{\prime}\right)
 =\int_{-\infty}^{\infty}\frac{{\text{d}} k}{2\pi}\phi\left(k\right)
    e^{ikz} \exp\left(-i\frac{E_{k}t^{\prime}}{\hbar}\right).
\end{equation}
So far this has been a standard analysis. Now, notice that $t^{\prime}$
is a function of $z$:
\begin{align}
 t^{\prime} &=t\left(1+\frac{gz}{c^{2}}\right),\label{transform}
 \\
\psi\left(z,t\right) & =\int_{-\infty}^{\infty}\frac{{\text{d}} k}{2\pi}\phi\left(k\right)e^{ikz}\exp\left[-i\frac{E_{k}t}{\hbar}\left(1+\frac{gz}{c^{2}}\right)\right]\\
 & =\int_{-\infty}^{\infty}\frac{{\text{d}}
   k}{2\pi}\phi\left(k\right)\exp\left[ iz\left(k-\frac{E_{k}gt}{\hbar
   c^{2}}\right)\right]
\exp\left(-i\frac{E_{k}t}{\hbar}\right).
\end{align}
We see that the wave packet is centered not around zero momentum but
around the time-dependent value
\begin{equation}
p\left(t\right)=\hbar k\left(t\right)=-\frac{E_{k}gt}{c^{2}}\simeq -mgt, \label{momentum}
\end{equation}
where we have approximated the energy by its rest value, $E_{k}\simeq mc^{2}$,
since the additional, $k$-dependent kinetic energy gives a correction
suppressed by inverse $c^{2}$ and is negligible for  non-relativistic
motion. 

The value of momentum in Eq.~\eqref{momentum} corresponds to the speed
\begin{equation}
v=gt,
\label{fromMom}
\end{equation}
as expected in the uniformly accelerated motion. 

\begin{figure}[h]
\centering\includegraphics[scale=0.4]{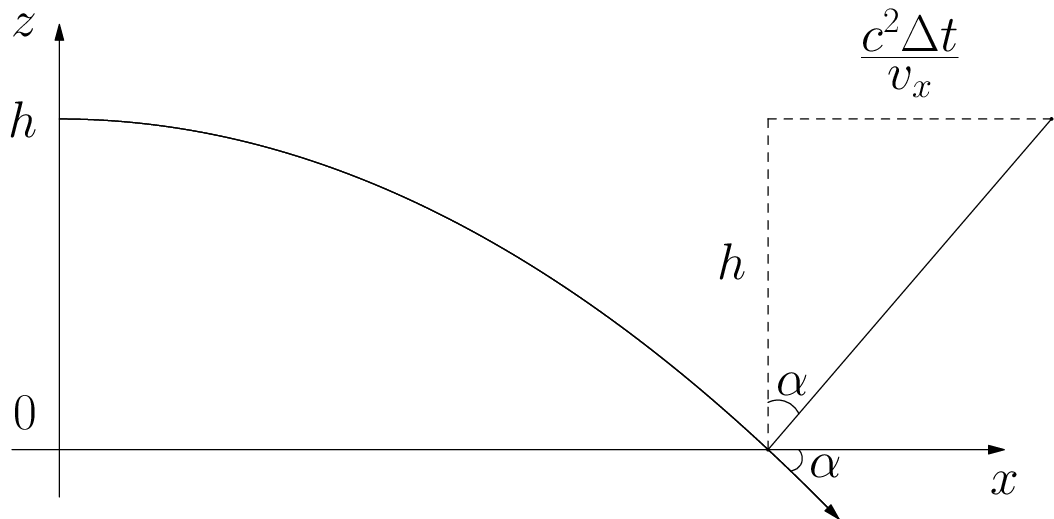}
\caption{Two-dimensional spatial trajectory of a freely falling 
  projectile (curved line). Similarly to Fig.~\ref{deBroglie}, the de 
  Broglie wave front (tilted solid line) changes direction, but here 
  it is shown in space rather than in spacetime, in order to  make it 
  easier to see the analogy with the refraction of ocean 
  waves.\label{free_fallX}}
\end{figure} 

We derive this result once more, in a manner that is less
abstract. Consider now a two-dimensional projectile motion, as shown
in Fig.~\ref{free_fallX}.
The projectile has a constant horizontal component of velocity
$v_x$. The phase velocity of de Broglie waves corresponding to this
motion is $c^2 / v_x$.\cite{Espinosa:1982aa} Thus the horizontal
side of the large triangle in Fig.~\ref{free_fallX} is  $c^2 \Delta t/ v_x$: an extra distance by
which the wave advances during the extra time $\Delta t$ elapsed at the
higher altitude. 
Using similar triangles, we relate the slope of the trajectory
to the sides of the large triangle,
\begin{equation}
- {\text{d} z \over \text{d} x} = \tan \alpha   = {1\over h}{c^2 \Delta t\over v_x}.
\end{equation}
On the other hand, $ \text{d} x = v_x  \text{d} t$, so that the
vertical component of the projectile's velocity is
\begin{equation}
- {\text{d} z \over \text{d} t} = {c^2 \Delta t\over h} = gt,
\end{equation}
as before in Eqs.~\eqref{eq:7} and \eqref{fromMom}.

\section{Jumping onto a train and the twin paradox}\label{other}

Here we show how Eq.~\eqref{DeltaT} helps to understand the twin
paradox. One twin stays at rest; the other sets out to travel with
a large velocity $v$. Each of them sees the other one moving and thus
each deduces that the sibling's  clock is running slow. Yet when the
twins reunite, the one who traveled turns out to have aged
less. Obviously, the symmetry is broken by the traveling twin having
to accelerate to reverse the direction of velocity and return. Yet it
may be hard to fathom that the extra aging of the twin at rest happens
only during that acceleration event. 

It is easier to consider a simpler, more localized situation: a
railway car of proper length $L$ is passing a station with speed
$v$. It is equipped with one clock at the front and one at the
rear. The clocks are synchronized in the car frame but from the point
of view of a ground observer, the rear clock is ahead by 
\begin{equation}
  \label{eq:5}
\Delta t = { L v \over c^2 };  
\end{equation}
see a lucid discussion on p.~513 in
Ref.~\onlinecite{morin2008Mech}.

Imagine that the ground observer decides to get on the train. After
sprinting in the direction of the train's motion (therefore towards the
front clock and away from the rear one), the observer sees both clocks
showing the same time. During the acceleration, the front clock must
have been running faster than the rear one. 

Denote the average acceleration of the sprinting observer by $g$, for
consistency with Eq.~\eqref{DeltaT}.  In order to reach the train's speed
$v$, the duration of the spurt is $t=v/g$.

\begin{figure}[h]
\centering\includegraphics[scale=0.5]{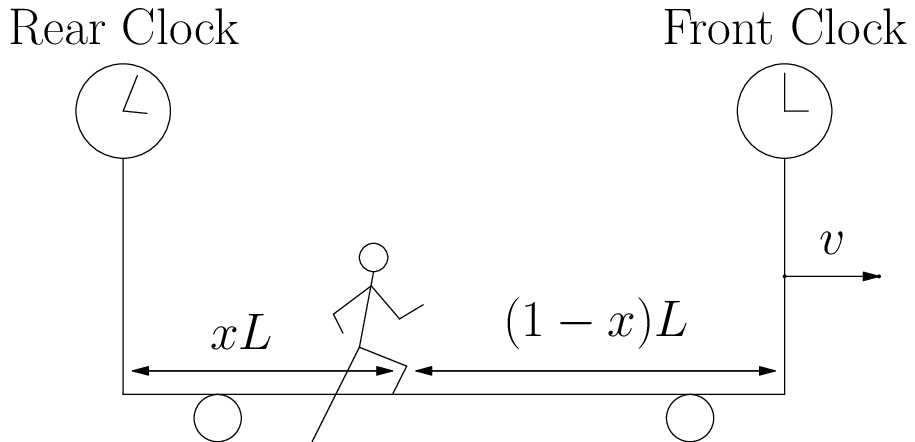}
\caption{ The rear clock on a moving train is ahead of the front 
  clock, when seen by a person standing on the ground.  In the frame 
  of an accelerating person jumping onto the train, the front clock 
  runs faster than the rear clock. For a person on the train, the 
  clocks show the same time.\label{fig:jump}}
\end{figure}
 Suppose the observer jumps on the train a distance $(1-x)L$ from the
front, $xL$ from the rear of the car, as shown in Figure
\ref{fig:jump}. According to Eq.~\eqref{DeltaT}, the front clock
registers an extra time
$\Delta t_F = g (1-x)Lt/c^2$ and the rear clock lags behind by
$\Delta t_R = g\,xL\,t/c^2$. The sum of these two effects gives the
net advance of the front clock,
\begin{equation}
  \label{eq:4}
  \Delta t = \Delta t_F + \Delta t_R = {gLt \over c^2 }  = {Lv \over c^2 }, 
\end{equation}
which exactly cancels the previous difference between the clocks,
Eq.~\eqref{eq:5}. The same mechanism resolves the twin paradox.

\section{Summary}\label{summary}

The gravitational time dilation effect on the Global Positioning
System (GPS)\cite{Mungan:2006aa,Ashby:2003aa} is often portrayed as
the most practically important effect of general relativity. We hope
that free fall will now inherit this distinction.  

Since the altitude of GPS satellites exceeds $20\,000$ km, it is not
suprising that gravitational effects are noticeably different on their
orbit than
on the earth's surface.  But how can the variation in the time flow be of
any relevance for phenomena near the surface? 
The effect of the terrestrial gravitational
field seems to be really small: one centimeter increase of the altitude,
$h_{1\,\text{cm}}$, causes a relative change of time flow of the order
of $\epsilon={g h_{1\,\text{cm}}} / {c^2} =10^{-18}$ (this is
approximately the current best precision of atomic
clocks.\cite{Takamoto:2020aa}) Yet not only is it relevant,
but it is the main mechanism causing bodies to fall.

The crucial reason is that the smallness of $\epsilon$, caused
by the inverse square of $c$, is compensated by the square of $c$
present in the rest energy $mc^2$. This cancellation leads to a $c$-independent
non-relativistic limit when $c\to \infty$.

From the point of view of a distant observer mentioned in the
Introduction, the de Broglie wave of the particle in a force field
evolves as $\exp -i\left( mc^2 + U\right)t$, where $U$ is the potential
energy. For example, gravitational potential energy  near Earth's
surface is $U=mgh$. The position
dependence of $U$ causes de Broglie waves to refract. This picture is
applicable to all  conservative forces. Gravity is unique in that one
can make a coordinate transformation, $\left( mc^2 + U\right)t =
mc^2 t^\prime $, see Eq.~\eqref{transform}, which is universal
because the $m$ dependence cancels. This universality leads to the geometrical
interpretation of gravity, as in
Ref.~\onlinecite{gould:2016aa}.\cite{Arkady}

We hope that the interpretation of the free fall in terms of de Broglie wave
refraction will help  students, especially those who do not have the time to
learn differential geometry, to grasp the relevance of general
relativity.

\section*{Note added}
After this paper was published, Robert Spekkens kindly informed us
about Ref.~\onlinecite{Wootters:2003up}, where the role of the gravitational
time dilation in the free fall had been pointed out.

\begin{acknowledgments}
  We thank Andrzej Bia{\l}as, Bruce A.~Campbell, Sacha Davidson,
  Valeri P.~Frolov, Frans R.~Klinkhamer, Chong-Sa Lim, Malcolm Longair, Don N.~Page,
  Andrzej Staruszkiewicz, and Arkady Vainshtein for helpful remarks.
  This work was supported by the Natural Sciences and Engineering
  Research Council of Canada.
\end{acknowledgments}

\appendix

\section{Gravitational time dilation: Einstein's derivation}
\label{AppEin}
Einstein discovered gravitational time dilation in 1907
(Ref.~\onlinecite{CollVol2Swiss00to09}, p. 301), long before he
created general relativity.  We note that it is common to all theories
that incorporate the Equivalence Principle.\cite{Will:2005yc,Sen:2020web} Here we
summarize Einstein's reasoning in deriving Eq.~\eqref{DeltaT}.

He started with the observation
that physical laws in a uniformly accelerated frame do not differ
from those in a frame at rest in a uniform gravitational field. 
Since he found an accelerated frame more theoretically accessible,
he used it to analyse the running of clocks and then inferred the corresponding
result in the gravitational field. 

He considered three reference frames: $S$ with spacetime coordinates $x,\,t$
is at rest; $\Sigma$ with coordinates $\xi,\,\sigma$ accelerates along
the $x$ axis with a constant acceleration $g$ with respect to an
instantaneously comoving inertial frame denoted by $S^{\prime}$,
with coordinates $x^{\prime},\,t^{\prime}$. The notion of a constant
acceleration was made precise in a later paper (Ref.~\onlinecite{CollVol2Swiss00to09},
p.~316), in response to a letter from Max Planck. 

At time $t=0$, $\Sigma$ is instantaneously at rest with respect
to $S$ and clocks everywhere in $\Sigma$ are set to $0$. The time
they measure is called the \emph{local time }in $\Sigma$ and is denoted
by $\sigma$. Local time at the origin of $\Sigma$, that is at $\xi=0$,
is  denoted by $\sigma(\xi=0)$.

Two events at different points $\xi$ are not in general simultaneous
with respect to the comoving frame $S^{\prime}$ when clocks at those
points show the same local time $\sigma$. Simultaneous events in
$S^{\prime}$ have the same value of $t^{\prime}$, related to coordinates
in $S$ by the Lorentz transformation,
\begin{equation}
t^{\prime}=t-\frac{vx}{c^{2}}.
\end{equation}
Time $\sigma(\xi=0)$ is considered so small that quadratic effects in $\sigma(\xi=0)$
and thus also in the velocity of $S^{\prime}$ and $\Sigma$ with
respect to $S$, $v=g\sigma(\xi=0)$, are neglected, thus the factor $\gamma=\left(1-{v^{2}}/{c^{2}}\right)^{{-1}/{2}}$
is approximated by 1.

Consider two events simultaneous in $S^{\prime}$ ($t_{1}^{\prime}=t_{2}^{\prime}\equiv t^{\prime}$):
one with coordinates $\left(x_{1},t_{1}\right)$, $\left(x_{1}^{\prime},t^{\prime}\right)$,
and $\left(\xi_{1},\sigma_{1}\right)$ respectively in $S$, $S^{\prime}$,
and $\Sigma$, and the other with subscripts $2$ instead of $1$.
The difference 
\begin{equation}
x_{2}-x_{1}=\left(x_{2}^{\prime}-vt^{\prime}\right)-\left(x_{1}^{\prime}-vt^{\prime}\right)=x_{2}^{\prime}-x_{1}^{\prime},
\end{equation}
is the same as $\xi_{2}-\xi_{1}$, since $\Sigma$ is at rest with
respect to $S^{\prime}$. Further, $t_{1}=\sigma_{1}$ and $t_{2}=\sigma_{2}$
because the duration of motion $\sigma(\xi=0)$ has been too short to destroy
the synchronization of the $\Sigma$ and $S$ clocks. Thus
\begin{align}
\sigma_{2}-\sigma_{1} & =t_{2}-t_{1}\\
 & =\left(t^{\prime}+\frac{vx_{2}}{c^{2}}\right)-\left(t^{\prime}+\frac{vx_{1}}{c^{2}}\right)\\
 & =\frac{v}{c^{2}}\left(x_{2}-x_{1}\right)=\frac{v}{c^{2}}\left(\xi_{2}-\xi_{1}\right).
\end{align}
Now suppose that event 1 takes place at the origin of $\Sigma$, $\xi_{1}=0$,
$\sigma_{1}=\sigma(\xi=0)$. Drop the subscript $2$ since the coordinates of event
2 now refer to any event in $\Sigma$:
\begin{equation}
\sigma-\sigma(\xi=0)=\frac{v}{c^{2}}\xi=\frac{g\sigma(\xi=0)}{c^{2}}\xi.
\end{equation}
Finally, 
\begin{equation}
\sigma=\sigma(\xi=0)\left(1+\frac{g\xi}{c^{2}}\right),
\end{equation}
equivalent to Eq.~\eqref{DeltaT}. 

\section{Wave equation in a uniform gravitational field}\label{waveeq}
Here we show that the non-relativistic limit of the Klein-Gordon
equation in the freely falling particle reference frame becomes a
Schr\"odinger equation with the gravitational potential $U(z) = mgz$
in the reference frame of a distant observer. We then apply the Ehrenfest
theorem\cite{SchiffQM} to reproduce the momentum evolution we found in
Eq.~\eqref{momentum}.  For a detailed study of the Klein-Gordon
equation and matter waves in a gravitational field see
Refs.~\onlinecite{Evans:1996aa,Evans:2001aa}.

In the particle reference frame,  the Klein-Gordon
equation is
\begin{equation}
  \label{eq:8}
-\partial_{t^{\prime}}^{2}\Psi(z,t^\prime)+c^{2}\nabla^{2}\Psi=\left(\frac{mc^{2}}{\hbar}\right)^{2}\Psi.
\end{equation}
Change the time variable to the time of the distant observer, see
Eq.~\eqref{transform}, 
$ t^\prime = \left( 1 + {gz\over c^2}\right)t$, so that $\partial_{t^{\prime}}^{2}\simeq\left(1-\frac{2gz}{c^{2}}\right)\partial_{t}^{2}$.
Factor out the leading time dependence, $
\Psi(z,t)=e^{-imc^{2}t/\hbar^{2}}\psi(z,t)$. Then
\begin{equation}
  \label{eq:9}
  \partial_{t}^{2}\Psi(z,t)=e^{-imc^{2}t/\hbar}\left[ \partial^2_t \psi-2\frac{imc^{2}}{\hbar}\partial_t \psi-\left(\frac{mc^{2}}{\hbar}\right)^{2}\psi\right],
\end{equation}
and the Klein-Gordon equation \eqref{eq:8} becomes
\begin{equation}
  \label{eq:10}
  \left[-\left(1-\frac{2gz}{c^{2}}\right)\partial^2_t \psi+2\left(1-\frac{2gz}{c^{2}}\right)\frac{imc^{2}}{\hbar}\partial_t \psi-\frac{2gz}{c^{2}}\left(\frac{mc^{2}}{\hbar}\right)^{2}\psi\right]+c^{2}\nabla^{2}\psi=0.
\end{equation}
Neglecting terms not enhanced by $c^{2}$ we obtain the Schr\"odinger
 equation with the gravitational potential $U(z) = mgz$,
 \begin{equation}
   \label{eq:11}
   i\hbar\partial_t \psi = -\frac{\hbar^{2}}{2m}\nabla^{2}\psi+mgz\psi.
 \end{equation}
The ground state solution of this equation and low-lying excitations have been tested in a series of
beautiful experiments with
neutrons.\cite{Nesvizhevsky:2002ef,Nesvizhevsky-Bounces} Here we are
interested in a rather high-energy behavior, where the gravitational
potential varies little over the size of a wave packet. In this limit the
Ehrenfest theorem is applicable. The evolution of the expectation
value of the momentum is given by
\begin{equation}
  \label{eq:12}
  {\text{d} \over \text{d}t} \langle p \rangle = \left\langle -
    {\partial U(z) \over \partial z} \right\rangle = -mg, 
\end{equation}
in agreement with Eq.~\eqref{momentum}.


\begin{thebibliography}{10}
\providecommand{\url}[1]{\texttt{#1}}
\providecommand{\urlprefix}{URL }
\providecommand{\eprint}[2][]{\url{#2}}

\bibitem{gould:2016aa}
R.~R. Gould, \emph{Why does a ball fall?: A new visualization for {Einstein's}
  model of gravity}, Am.~J.~Phys. \textbf{84}, 396 (2016).

\bibitem{greene2008icarus}
B.~Greene, \emph{Icarus at the Edge of Time}, Alfred A. Knopf, New York (2008).

\bibitem{Rebilas2017}
K.~R\k{e}bilas, \emph{Comment on ``Why does a ball fall?: A new visualization
  for Einstein's model of gravity'' [Am. J. Phys. 84, 396--402 (2016)]},
  Am.~J.~Phys. \textbf{85}, 66--67 (2017).

\bibitem{Dieks:1990aa}
D.~Dieks and G.~Nienhuis, \emph{Relativistic aspects of nonrelativistic quantum
  mechanics}, Am.~J.~Phys. \textbf{58}, 650--655 (1990).

\bibitem{Houchmandzadeh:2020aa}
B.~Houchmandzadeh, \emph{The Hamilton--Jacobi equation: An alternative
  approach}, Am.~J.~Phys. \textbf{88}, 353--359 (2020).

\bibitem{Stannard_2016}
W.~Stannard, D.~Blair, M.~Zadnik, and T.~Kaur, \emph{Why did the apple fall? A
  new model to explain Einstein's gravity}, Eur.~J.~Phys. \textbf{38}, 015603
  (2016).

\bibitem{AniaFilm}
A.~Czarnecka, \emph{Gravitational Time Dilation} (2020),
  https://youtu.be/Hn6eKbexUWA.

\bibitem{Takamoto:2020aa}
M.~Takamoto, I.~Ushijima, N.~Ohmae, T.~Yahagi, K.~Kokado, H.~Shinkai, and
  H.~Katori, \emph{Test of general relativity by a pair of transportable
  optical lattice clocks}, Nature Photonics \textbf{14}, 411--415 (2020).

\bibitem{Hafele166}
J.~C. Hafele and R.~E. Keating, \emph{Around-the-World Atomic Clocks: Predicted
  Relativistic Time Gains}, Science \textbf{177}, 166--168 (1972).

\bibitem{Hafele168}
J.~C. Hafele and R.~E. Keating, \emph{Around-the-World Atomic Clocks: Observed
  Relativistic Time Gains}, Science \textbf{177}, 168--170 (1972).

\bibitem{GravityProbeA}
R.~F.~C. Vessot, M.~W. Levine, E.~M. Mattison, E.~L. Blomberg, T.~E. Hoffman,
  G.~U. Nystrom, B.~F. Farrel, R.~Decher, P.~B. Eby, C.~R. Baugher, J.~W.
  Watts, D.~L. Teuber, and F.~D. Wills, \emph{Test of Relativistic Gravitation
  with a Space-Borne Hydrogen Maser}, Phys. Rev. Lett. \textbf{45}, 2081--2084
  (1980).

\bibitem{Delva_2018}
P.~Delva, N.~Puchades, E.~Sch{\"o}nemann, F.~Dilssner, C.~Courde, S.~Bertone,
  F.~Gonzalez, A.~Hees, C.~Le~Poncin-Lafitte, F.~Meynadier, and et~al.,
  \emph{Gravitational Redshift Test Using Eccentric {Galileo} Satellites},
  Phys.~Rev.~Lett. \textbf{121}, 231101 (2018).

\bibitem{Herrmann_2018}
S.~Herrmann, F.~Finke, M.~L{\"u}lf, O.~Kichakova, D.~Puetzfeld, D.~Knickmann,
  M.~List, B.~Rievers, G.~Giorgi, C.~G{\"u}nther, and et~al., \emph{Test of the
  Gravitational Redshift with {Galileo} Satellites in an Eccentric Orbit},
  Phys.~Rev.~Lett. \textbf{121}, 231102 (2018).

\bibitem{Van-Baak:2007aa}
T.~{Van Baak}, \emph{An adventure in relative time-keeping}, Physics Today
  \textbf{60}, 16 (2007), {see also http://leapsecond.com/great2005}.

\bibitem{Espinosa:1982aa}
J.~M. Espinosa, \emph{Physical properties of de Broglie's phase waves},
  American Journal of Physics \textbf{50}, 357--362 (1982).

\bibitem{morin2008Mech}
D.~Morin, \emph{Introduction to Classical Mechanics: With Problems and
  Solutions}, Cambridge University Press, Cambridge (2008).

\bibitem{Mungan:2006aa}
C.~E. Mungan, \emph{{Relativistic Effects on Clocks Aboard GPS Satellites}},
  The Physics Teacher \textbf{44}, 424--425 (2006).

\bibitem{Ashby:2003aa}
N.~Ashby, \emph{Relativity in the Global Positioning System}, Living Reviews in
  Relativity \textbf{6}, 1 (2003).

\bibitem{Arkady}
We thank Arkady Vainshtein for this interpretation. 

\bibitem{Wootters:2003up}
W.~K. Wootters, \emph{Why Things Fall}, Foundations of Physics \textbf{33},
  1549--1557 (2003).

\bibitem{CollVol2Swiss00to09}
J.~Stachel, D.~C. Cassidy, J.~Renn, and R.~Schulmann, editors, \emph{The
  Collected Papers of Albert Einstein. Vol. 2: The Swiss Years: Writings,
  1900-1909 (English translation)}, Princeton University Press (1989).

\bibitem{Will:2005yc}
C.~M. Will, \emph{{Was Einstein right?}: {testing relativity at the
  centenary}}, Annalen Phys. \textbf{15}, 19--33 (2005),
  \eprint{gr-qc/0504086}.

\bibitem{Sen:2020web}
A.~Sen, S.~Dhasmana, and Z.~K. Silagadze, \emph{{Free fall in KvN mechanics and
  Einstein's principle of equivalence}}, Annals Phys. \textbf{422}, 168302
  (2020), \eprint{2009.02914}.

\bibitem{SchiffQM}
L.~I. Schiff, \emph{Quantum Mechanics}, McGraw-Hill, New York, 3rd edition
  (1968).

\bibitem{Evans:1996aa}
J.~Evans, K.~K. Nandi, and A.~Islam, \emph{The optical--mechanical analogy in
  general relativity: New methods for the paths of light and of the planets},
  Am.~J.~Phys. \textbf{64}, 1404--1415 (1996).

\bibitem{Evans:2001aa}
J.~Evans, P.~M. Alsing, S.~Giorgetti, and K.~K. Nandi, \emph{Matter waves in a
  gravitational field: An index of refraction for massive particles in general
  relativity}, Am.~J.~Phys. \textbf{69}, 1103--1110 (2001).

\bibitem{Nesvizhevsky:2002ef}
V.~V. Nesvizhevsky et~al., \emph{{Quantum states of neutrons in the Earth's
  gravitational field}}, Nature \textbf{415}, 297--299 (2002).

\bibitem{Nesvizhevsky-Bounces}
V.~Nesvizhevsky and A.~Voronin, \emph{Surprising Quantum Bounces}, Imperial
  College Press (2015).

\end{thebibliography}

\end{document}